\documentstyle[aps,prl,multicol,floats,epsf]{revtex}

\newcommand{\be}{\begin{equation}}
\newcommand{\ee}{\end{equation}}
\def\(#1){(\ref{#1})}

\newcommand{\eg}{{\em e.g.}}
\newcommand{\ie}{{\em i.e.}}

\newcommand{\rh}{\rho}
\newcommand{\sig}{\sigma}
\newcommand{\rhosig}{\rh(\sig)}
\newcommand{\rhonsig}{\rh\pn(\sig)}
\newcommand{\prior}{R(\sig)}
\newcommand{\musig}{\mu(\sig)}
\newcommand{\intsig}{\int\!d\sig\,}

\newcommand{\wi}{w_i(\sig)}
\newcommand{\m}{m}
\newcommand{\mi}{m_i}

\newcommand{\mo}{m_1}
\newcommand{\mhat}{\lambda}

\newcommand{\mhati}{\mhat_i}

\newcommand{\fexc}{\tilde{f}}

\newcommand{\feff}{f_{\rm pr}}
\newcommand{\shat}{m_0}
\newcommand{\al}{\alpha}
\newcommand{\rhoalsig}{\rh\pa(\sig)}
\newcommand{\phal}{\phi\pa}
\newcommand{\pn}{^{(0)}}

\newcommand{\pa}{^{(\al)}}

\title{Projected free energies for polydisperse phase equilibria}

\author{Peter Sollich$^*$ and Michael E. Cates}

\address{Department of Physics and Astronomy, University of Edinburgh,
Edinburgh EH9 3JZ, U.K.}

\draft 
\begin{document}

\maketitle

\begin{abstract}
A `polydisperse' system has an infinite number of conserved
densities. We give a rational procedure for projecting its
infinite-dimensional free energy surface onto a subspace comprising a
finite number of linear combinations of densities (`moments'), in
which the phase behavior is then found as usual. If the {\em excess}
free energy of the system depends only on the moments used, exact
cloud, shadow and spinodal curves result; two- and multi-phase regions
are approximate, but refinable indefinitely by adding extra
moments. The approach is computationally robust and gives new
geometrical insights into the thermodynamics of polydispersity.
\end{abstract}

\pacs{PACS numbers: 05.20.-y, 64.10.+h, 82.70.-y, 61.25.Hq. To appear
in {\em Physical Review Letters}.}

\begin{multicols}{2}

The thermodynamics of mixtures of several chemical species is, since
Gibbs, a well established subject (see \eg~\cite{DeHoff92}). But many
systems arising in nature and in industry contain, for practical
purposes, an infinite number of distinct, though similar, chemical
species. Often these can be classified by a parameter, $\sigma$, say,
which could be the chain length in a polymeric system, or the particle
size in a colloid; both are routinely treated as continuous
variables. In other cases (see
\eg~\cite{Bauer85,RatWoh91,NesOlvCri93,Solc93}) $\sigma$ is instead a
parameter distinguishing species of continuously varying chemical
properties.  The thermodynamics of polydispersity (thus defined) is
therefore of crucial interest to wide areas of science and technology.

Standard thermodynamic procedures~\cite{DeHoff92} for constructing
phase equilibria in a system of volume $V$ containing $M$ different
species can be understood geometrically in terms of a free energy
surface $f(\rho_j)$ (with $f=F/V$) in the $M$-dimensional space of
density variables $\rho_j$. Tangent planes to $f$ define regions of
coexistence, within which the free energy of the system is lowered by
phase separation. The volumes of coexisting phases follow from the
well-known `lever rule'\cite{DeHoff92}. Here `surface' and `plane' are
used loosely, to denote manifolds of appropriate dimension. This
procedure becomes unmanageable, both conceptually and numerically, in
the limit ($M\to\infty$) of a polydisperse system.  There is now a
separate conserved density $\rho(\sigma)$ for each value of $\sigma$;
the overall density of particles is $\rh=\intsig \rhosig$.  The free
energy surface is $f = f[\rho(\sigma)]$ which resides in an infinite
dimensional space. Gibbs' rule allows the coexistence of arbitrarily
many thermodynamic phases.

Experimentally, one often restricts attention to the cloud- and
shadow-curves (also referred to as dew/bubble curves). For a fixed
`shape' of polydispersity $\tilde\rho(\sigma) = \rho(\sigma)/\rho$,
these define as a function of the overall particle density $\rho$ (and
temperature $T$) the onset of two-phase equilibrium (cloud-curve) and
the density of the corresponding minority phase (its shadow); see
\eg~\cite{RatWoh91,IrvKen82,ClaMcLeiJen95}.  Theoretical work has
likewise focused on attempts to bring the problem into a more
manageable form by somehow reducing its
dimensionality~\cite{RatWoh91,NesOlvCri93,Solc93,IrvKen82,%
SalSte82,SolKon95,GuaKinMor82,IrvGor81,BeeBerKehRat86,%
Hendriks88,HenVan92,CotPra85,ShiSanBeh87,Bartlett97}.  In this Letter
we propose a new and general method, whereby we construct from
$f[\rho(\sigma)]$ an (optimally) projected free energy surface in a
reduced subspace of density variables. For these variables, we choose
linear combinations of densities, the `generalized moments'
$\mi=\intsig \wi \rhosig$ of $\rhosig$, defined by certain weight
functions $\wi$; these are ordinary (non-normalized) moments if
$\wi=\sig^i$.

The simplest imaginable case is where the free energy $f$ depends only
on a finite set of $K$ such moments:
\be
f = f(m_i), \qquad i=1\ldots K \label{puremoment}
\ee
In coexisting phases one demands equality of particle chemical
potentials, defined as $ \musig={\delta f/\delta\rhosig}=\sum_i
({\partial f/\partial \mi}) \wi=\sum_i \mu_i \wi $, for all
$\sig$. But this implies that all `moment' chemical potentials,
$\mu_i\equiv{\partial f}/{\partial \mi}$, are likewise equal among
phases. The osmotic pressures $\Pi$ of all phases also must be equal;
simple algebra establishes that $ -\Pi = f-\sum_i\mu_i\mi$ which
involves only the moments $m_i$ and their chemical potentials $\mu_i$.
Finally, if the overall $\sigma$-distribution is $\rhonsig$, and there
are $p$ coexisting phases with $\sigma$-distributions $\rhoalsig$,
each occupying a fraction $\phal$ of the total volume ($\al=1\ldots
p$), then conservation of particles implies the usual `lever rule' (or
material balance) among species: $ \sum_\al \phal \rhoalsig =
\rhonsig, \forall \sigma$.  Multiplying this by a weight function
$\wi$ and integrating over $\sig$ shows that the lever rule also holds
for the moments:
\be 
\sum_{\al=1}^p \phal \mi\pa = \mi\pn
\label{lever_moments}
\ee
These results express the fact that any linear combination of
conserved densities (a generalized moment) is itself a conserved
density in thermodynamics. Therefore, if the free energy of the system
depends only on $K$ moments $\mi\ldots m_K$ we can view these as the
densities of $K$ `quasi-species' of particles, and construct the phase
diagram via the usual construction of tangencies and the lever rule.
Formally this has reduced the problem to finite dimensionality by a
projection, although this is trivial here because $f$, by
construction, has no dependence on any variables other than the $m_i$
($i=1\ldots K$).
\end{multicols}
\twocolumn

Of course, it is uncommon for the free energy $f$ to
obey~\(puremoment). In particular, the `ideal gas' (or, for polymers,
Flory-Huggins) entropy term, in mixtures of many species, is
definitely not of this form. On the other hand, in very many
thermodynamic (especially mean field) models the free energy takes the
form ($k_{\rm B}=1$)
\be
f = \fexc(\mi) + T \intsig \rhosig \left[\ln \left(\rhosig/R(\sigma)\right)
-1\right],
\label{free_en_decomp}
\ee
in which the {\em excess} free energy $\fexc$ {\em does} depend only
on $K$ moments. Examples include polydisperse hard
spheres~\cite{SalSte82}, polydisperse homo- and
copolymers~\cite{Bauer85,RatWoh91,NesOlvCri93,Solc93,SolKon95}, and
van der Waals fluids with factorized interaction
parameters~\cite{GuaKinMor82}.  Note that, in the ideal gas term of
(\ref{free_en_decomp}), we have included a dimensional factor
$R(\sigma)$ inside the logarithm: since the resulting contribution is
linear in densities, this has no effect in rigorous
thermodynamics. However, it will play a central role in our approach.

In principle, the phase equilibria stemming from
(\ref{free_en_decomp}) can be computed exactly by a finite
algorithm. Specifically, the spinodal stability criterion involves a
$K$-dimensional square
matrix~\cite{IrvGor81,BeeBerKehRat86,Hendriks88,HenVan92} whereas
calculation of $p$-phase equilibrium involves solution of $(p-1)(K+1)$
strongly coupled nonlinear equations. This method has certainly proved
useful~\cite{Bauer85,RatWoh91,NesOlvCri93,SolKon95,GuaKinMor82,HenVan92},
but is cumbersome, particularly if one is interested mainly in cloud-
and shadow-curves, rather than coexisting compositions deep within
multi-phase
regions~\cite{Bauer85,NesOlvCri93,ClaMcLeiJen95,SolKon95}. Various
ways of simplifying the procedure
exist~\cite{IrvKen82,IrvGor81,CotPra85,ShiSanBeh87,Bartlett97}, but
there has been, up to now, no systematic alternative to the full
computation. Note also that the nonlinear phase equilibrium equations
permit no simple geometrical interpretation or qualitative insight
akin to the familar rules for constructing phase diagrams from the
free energy surface of a finite mixture.

Our method instead proceeds by deriving from (\ref{free_en_decomp}) a
`projected' free energy that depends only on a finite set of
moments. We argue that the most important moments to treat correctly
are those that actually appear in the excess free energy $\fexc(m_i)$.
Accordingly we divide the infinite-dimensional space of
$\sigma$-distributions into two orthogonal subspaces: a `moment
subspace', which contains all the degrees of freedom of $\rhosig$ that
contribute to the moments $\mi$ (this subspace is spanned by the
weight functions $\wi$), and a `transverse subspace' which contains
all remaining degrees of freedom (as can be varied without affecting
the chosen moments $m_i$).  Physically, it is reasonable to expect
that these `leftover' degrees of freedom play a relatively minor role
in the phase equilibria of the system, a view justified {\em a
posteriori} below. Accordingly, we now allow violations of the lever
rule, so long as these occur {\em solely in the transverse space}.
The `transverse' degrees of freedom, instead of obeying the strict
particle conservation laws, are chosen so as to minimize the free
energy: they are treated as `annealed'.  If, as assumed above,
$\fexc=\fexc(\mi)$ only depends on the moments retained, this amounts
to {\em maximizing the entropy} in (\ref{free_en_decomp}), while
holding fixed the values of the moments $m_i$.

At this point, the factor $\prior$ in (\ref{free_en_decomp}), which is
immaterial if all conservation laws are strictly obeyed, becomes
central. Indeed, maximizing the entropy over all distributions
$\rhosig$ at fixed moments $\mi$ yields
\be
\rhosig=\prior\exp\left(\sum_i \mhati\wi\right)
\label{family}
\ee
where the Lagrange multipliers $\mhati$ are chosen to give
\be
\mi = \intsig\wi\,\prior\exp\left(\sum_i \mhati\wi\right)
\label{moments_from_lambda}
\ee
The corresponding minimum value of $f$ then defines our projected
(\ie, annealed) free energy
\be
\feff(\mi) = \fexc + T \left(\sum_i \mhati \mi - \shat\right) \label{annd}
\ee
In the last term, $m_0 = \intsig \rhosig$ is the `zeroth moment' which
is identical to the overall particle density $\rho$ defined
previously. If this is among the moments used for the projection, the
resulting linear term can be dropped; otherwise it must be retained
(with $\shat$ now expressed as a function of the $\mi$, via the
$\mhati$).

Our maximum entropy method yields a free energy $\feff(\mi)$ which
only depends on the chosen set of moments: \ie,~\(annd) is of the form
(\ref{puremoment})~\cite{lambda_vs_m_footnote}. A finite dimensional
phase diagram can now be constructed from it according to the usual
rules.  Obviously, though, the results now depend on $\prior$ which is
formally a `prior distribution' for the entropy maximization. To
understand its thermodynamic role, we recall that our projected free
energy $\feff(m_i)$ was constructed as the minimum of
$f[\rho(\sigma)]$ at fixed $\mi$; that is, $\feff$ is the lower
envelope of the projection of $f$ onto the moment subspace.
Crucially, the shape of this envelope depends on how, by choosing a
particular prior distribution $\prior$, we `tilt' the
infinite-dimensional free energy surface {\em before} projecting it.

To find the optimum choice of prior, we note that $\prior$ serves
physically to determine which distributions $\rhosig$ lie within the
maximum-entropy family~\(family) that the annealed system can
have. Typically, one is interested in a system where a fixed overall
`parent' (or `feed') distribution $\rhonsig$ becomes subject to
separation into various phases. In such circumstances, we should
generally choose this parent distribution as our prior,
$\prior=\rhonsig$, thereby guaranteeing that it is contained within
the family~\(family). Having done this, we note that the annealing
procedure will be {\em exactly valid}, to whatever extent the
$\sigma$-distributions {\em actually arising} in the various
coexisting phases of the system under study {\em are members of the
family}~\(family).  (This statement of exactness, and similar ones
below, of course hold only if~\(free_en_decomp) is valid.)

In fact, the condition just described does hold whenever all but one
of a set of coexisting phases are of infinitesimal volume compared to
the majority phase. This is because the $\sigma$-distribution,
$\rhonsig$, of the majority phase is negligibly perturbed, whereas
that in each minority phase differs from this by an exponential
Gibbs-Boltzmann factor, of exactly the form required for~\(family).
Accordingly, our projection method yields {\em exact} cloud-curves and
shadow-curves. By the same argument, critical points (which in fact
lie at the intersection of these two curves) are exactly
determined. Moreover, all spinodals are also found exactly by our
annealing method. For, at a spinodal, there exists an instability
direction (in the full space) along which the curvature of the free
energy vanishes; in all other directions $f$ has positive
curvature. One can show that such an instability direction always
connects neighboring distributions within the same maximum entropy
family~\(family), and hence that only the free energy of such
distributions (\ie, the projected free energy with the parental
$R(\sigma)$) is needed to calculate spinodals. The geometrical
interpretation of this result, and also proofs of it and the others
stated above, will be given elsewhere~\cite{polydisp_long}.

The method does, however, give only approximate results for
coexistences involving finite amounts of different phases. This is
because linear combinations of different $\sigma$-distributions
obeying~\(family), corresponding to two (or more) phases arising from
the same parent ($\rhonsig=\prior$) do not necessarily add to recover
the parent distribution itself. Moreover, according to Gibbs' phase
rule, a projected free energy depending on $n$ moments will not
normally predict more than $n+1$ coexisting phases, whereas a
polydisperse system can in principle separate into an arbitrary number
of phases. Both of these shortcomings can be overcome by
systematically including additional moments within the annealing
procedure.  (The above exact results are unaffected, because these do
not exclude a null dependence of $\fexc$ on certain of the $\mi$.)
Indeed, by adding further moments one can indefinitely expand the
maximum-entropy family~\(family) of $\sigma$-distributions, thereby
approaching with increasing precision the actual distributions in all
phases present; this yields phase diagrams of ever-refined
accuracy. How quickly convergence to the exact results occurs depends
on the choice of weights functions for the additional moments; this
will be quantified elsewhere~\cite{polydisp_long}.

To demonstrate the power of our approach, we consider a specific
example.  This is a simplified model of chemical fractionation, in
which one considers species of continuously variable chemical
character (such as aromaticity) governed by a parameter $\sig$ between
0 and 1. We suppose that the interaction energy between species varies
as $(\sig-\sig')^2$, so that the most different species repel each
other most strongly. For simplicity we take a molten system, choosing
volume units so that the overall density is constrained as
$\intsig\rhosig=1$.  Within a mean-field treatment, the system is then
described by a free energy of the form~\(free_en_decomp), with an
excess free energy (in units of $k_{\rm B}T$) of $\fexc=-\chi\mo^2$,
$\mo=\intsig \sig \rhosig$ (up to irrelevant terms linear in
$\rhosig$).  This model differs only by a rescaling of parameters
(with powers of polymer molecular weight) from the Flory-Huggins
treatment of random AB copolymers, in which $\chi$ is the usual
interaction parameter and $\sigma$ is the proportion of A monomers in
a chain~\cite{Bauer85,RatWoh91,NesOlvCri93}.  The model should show
fractionation into an ever-increasing number of phases as $\chi$ is
increased.  It is therefore an interesting test case for our
projection approach (and the method of adding further moments), yet
simple enough for exact phase equilibrium calculations to remain
feasible, allowing detailed comparisons to be made.

\begin{figure}
\vspace*{-3.3mm}
\begin{center}
\leavevmode
\epsfxsize 240pt
\epsfbox{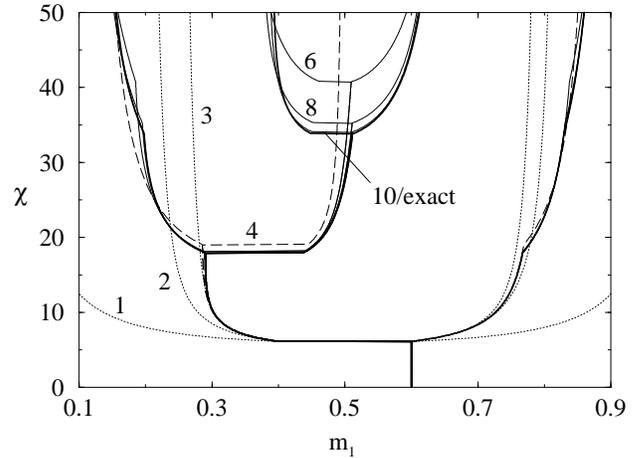}
\end{center}
\vspace*{-0.2cm}
\caption{Coexistence curves for a parent distribution with
$\mo\pn=0.6$. Shown are the values of $\mo$ of the coexisting phases;
horizontal lines guide the eye where new phases appear. Curves are
labeled by $n$, the number of moments retained in the projected free
energy. Predictions for $n=10$ are indistinguishable from an exact
calculation (in bold).
\label{fig:comparison}
}
\end{figure}

We consider phase separation from parent phases with $\sigma$
distributions of the form $\rhosig\propto\exp(\lambda\sig)$ (for
$0\le\sig\le 1$); $\lambda$ is thereby fixed in terms of the parental
$\mo=\mo\pn$.  Fig~\ref{fig:comparison} shows the exact coexistence
curve for $\mo\pn=0.6$, along with the predictions from our projected
free energy with $n$ moments ($\m_i=\intsig\sig^i\rhosig$, $i=1\ldots
n$) retained. Comparable results are found for other $\mo\pn$. Even
for the minimal set of moments ($n=1$) the point where phase
separation first occurs on increasing $\chi$ is predicted correctly
(this is a cloud point for the given parent). As more moments are
added, the annealed coexistence curves approach the exact one to
higher and higher precision~\cite{precision_footnote}. As expected,
the precision decreases at high $\chi$, where fractionated phases
proliferate; in this region, the number of coexisting phases predicted
by the projection method increases with $n$.  However, it is not
always equal to $n+1$, as one might expect from a naive use of Gibbs'
phase rule; three-phase coexistence, for example, is first predicted
for $n=4$~\cite{phase_rule_footnote}. Note that the stability of the
results to the addition of extra moments provides, in this example, a
good test of convergence on the coexistence curves.

For the computational implementation of both the annealing and the
exact method we used a Newton-Raphson nonlinear equation solver. The
annealed calculation turned out to be significantly more robust with
respect to the choice of initial values, the size of $\chi$ increments
etc. due to an effective decoupling of the equations: Equality of
chemical potentials is achieved using the moments contained in the
excess free energy, while the lever rule is satisfied (increasingly
accurately) using the remaining moments~\cite{polydisp_long}. This
advantage should be much more pronounced in more complex cases, as
should savings in computer time (which are modest in our simple
example).

With exact results for cloud- and shadow-curves, critical points and
spinodals, as well as refinably accurate coexistence curves and
multi-phase regions, our annealing method allows rapid and accurate
computation of the phase behavior of many polydisperse
systems. Moreover, by establishing the link to a projected free energy
$\feff(m_i)$ as a function of a finite set of conserved densities
$m_i$, it restores to the problem much of the geometrical
interpretation and insight (as well as the computational methodology)
associated with phase diagrams for finite mixtures.  This contrasts
with procedures commonly used for systems in which the excess free
energy involves a finite set of moments~\(free_en_decomp)
\cite{Bauer85,RatWoh91,NesOlvCri93,Solc93,SolKon95,GuaKinMor82}.  Some
previous approximations to that problem have used (generalized)
moments as coordinates; see
\eg~\cite{IrvGor81,HenVan92,CotPra85,Bartlett97,binning}. Our
annealing method provides a rational basis for these methods and, by a
careful choice of prior, guarantees that many properties of interest
are found exactly.

Finally, our method may extend to models for which the excess free
energy {\em cannot} be written directly in terms of a finite number of
moments as in~\(free_en_decomp). For example, many mean-field theories
correspond to a variational minimization of the free energy: $F\le
\langle E\rangle_0 - TS_0$, where subscript $0$ refers to a trial
Hamiltonian~\cite{Feynman72}. In such a case, one might choose to {\em
first} make a physically motivated decision about which (and how many)
moments $\mi$ to keep, and then include among the variational
parameters the annealed ``transverse" degrees of freedom. This would
lead directly to a mean-field estimate of the projected free energy
without assuming Eq.~\(free_en_decomp). Note that a good choice of
prior $R(\sigma)$ will again be important. Although no exact results
can be guaranteed, this approach may form a promising basis for future
developments.

{\em Acknowledgements:} After this work was substantially complete, we
learned from P. B. Warren~\cite{Warren97} that he has independently
developed an approach which, though based on distinctly different
principles, yields a formalism broadly equivalent to our
own~\cite{polydisp_long}. We thank him, and also N. Clarke,
R. M. L. Evans, T. McLeish, P. Olmsted, and W. C. K. Poon, for helpful
discussions.


\begin{thebibliography}{10}
\bibitem{lambda_vs_m_footnote}
Numerically, it is easier to view the annealed free energy as a
function of the Lagrange multipliers $\mhati$ rather than the moments
$\mi$, to avoid having to invert
eqs.~(\protect\ref{moments_from_lambda}).
\bibitem{phase_rule_footnote} 
For fractionation problems such as this (but not more generally), the
low temperature limit suggests that to obtain $n+1$ phases, one may
have to include up to $2n$ moments.  Using localized weight functions
(rather than powers of $\sigma$) for the extra moments can reduce this
number, but gives less accurate coexistence
curves~\protect\cite{polydisp_long}.
\bibitem{binning}
Note also that the commonplace method of `binning' the
$\sig$-distribution into discrete `pseudo-components' is a special
case of our annealing method in which each weight function is zero
outside the corresponding bin.
\end{thebibliography}
\end{document}